\documentclass[twocolumn,showpacs,preprintnumbers,amsmath,amssymb]{revtex4}

\usepackage{graphicx}% Include figure files
\usepackage{dcolumn}% Align table columns on decimal point
\usepackage{bm}% bold math

\usepackage{amsmath}
\usepackage{multirow}
\usepackage{amssymb}

\begin{document}

%\preprint{APS/123-QED}
\title{Dynamical analysis of the exclusive queueing process}

\author{Chikashi Arita}
\email{arita@math.kyushu-u.ac.jp}
\affiliation{
Faculty of Mathematics, Kyushu University, Fukuoka, 819-0395, Japan}

\author{Andreas Schadschneider}
\email{as@thp.uni-koeln.de}
\affiliation{
Institute for Theoretical Physics,
University of Cologne, 50937 Cologne, Germany}
%\date{\today}

\newcommand{\Jout}{{ J^{\rm out} }}
\newcommand{\twotwo}
[4]{\left(\begin{array}{cc} #1 & #2 \\ #3 & #4\end{array}\right)}
\newcommand{\cZ}{{\mathcal Z}}
\newcommand{\cY}{{\mathcal Y}}
\newcommand{\cX}{{\mathcal X}}
\newcommand{\W}{\langle W |}
\newcommand{\V}{| V \rangle}
\newcommand{\cA}{{\mathcal A}}
\newcommand{\cC}{{\mathcal C}}
\newcommand{\cT}{{\mathcal T}}
\newcommand{\bbZ}{{\mathbb Z}}
\newcommand{\bbN}{{\mathbb N}}
\newcommand{\factor}{\left(\frac{\alpha}{p(1-\alpha)}\right)}

%%%%%%%%%%%%%%%%%%%%%%%%%%%%%%%%%%%%%%%%%%%%%%%%%%%%%%%%%%%%%%%%%%
%%%%%%%%%%%%%%%%%%%%%%%%%%%%%%%%%%%%%%%%%%%%%%%%%%%%%%%%%%%%%%%%%%
%%%%%%%%%%%%%%%%%%%%%%%%%%%%%%%%%%%%%%%%%%%%%%%%%%%%%%%%%%%%%%%%%%

%\titlerunning{Short form of title}% if too long for running head

\begin{abstract}
  In a recent study \cite{RefAY} the stationary state of a
  parallel-update TASEP with varying system length, which can be
  regarded as a queueing process with excluded-volume effect ({\it
    exclusive queueing process}, EQP), was obtained.  We analyze the
  dynamical properties of the number of particles $\langle N_t\rangle$
  and the position of the last particle (the system length) $\langle
  L_t\rangle$, using an analytical method (generating function technique)
  as well as a phenomenological description based on domain wall
  dynamics and Monte Carlo simulations. The system exhibits two phases
  corresponding to linear convergence or divergence of $\langle
  N_t\rangle$ and $\langle L_t\rangle$. These phases can both further
  be subdivided into high-density and maximal-current subphases.  
  The predictions of the domain wall theory are found to be in very
  good agreement quantitively with results from Monte Carlo
  simulations in the convergent phase.  On the other hand, in the
  divergent phase, only the prediction for $\langle N_t\rangle$ agrees
  with simulations.  
\end{abstract}



\maketitle

\begin{figure}\begin{center}
\centering
\includegraphics{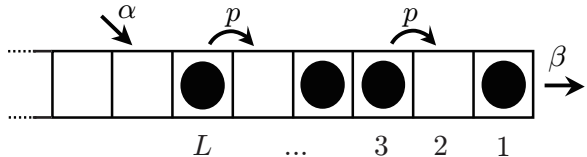}
\caption{Exclusive queueing process.}
\label{fig:eq} 
\end{center}\end{figure}

\section{Introduction}\label{intro}

Queueing processes have been studied extensively, especially due to
their practical relevance \cite{RefE,RefK,RefS}.  However, usually the
spatial structure of the queues is neglected and the particles in the
queues do not interact with each other.  
On the other hand, the totally asymmetric simple exclusion process
(TASEP) which has a spatial structure and excluded-volume effect
(hard-core repulsion) is one of the best-studied
interacting particle systems \cite{RefL}.  Nowadays the TASEP is a
basic model for pedestrian and traffic flows \cite{RefCSS,RefSCN}.

Recently a queueing process with the excluded-volume effect,
the exclusive queueuing process (EQP),  
was introduced in \cite{RefA} and \cite{RefY} independently, where the
model was formulated as continuous-time and discrete-time Markov
processes, respectively.
This model can be rephrased as the TASEP on a semi-infinite chain
with a new boundary condition, see Fig.~\ref{fig:eq}. 
The left end is interpreted as the end of the queue where new
customers arrive. Therefore  particles can enter at the left site next to
the leftmost occupied site.
The (fixed) rightmost site corresponds to the server.
Here particles can leave the system after getting service.
In the bulk a particle can hop to its right nearest neighbor site,
if the target site is empty. 
After the study \cite{RefY}, where the bulk hopping rule is
deterministic, the model with discrete time and probabilistic bulk
hopping was analyzed \cite{RefAY}.  

Earlier works \cite{RefA,RefAY,RefY} focussed on the exact
probability distribution, physical quantities in the stationary state,
and the conditions under which the EQP converges to the stationary
state.  In this paper, we study the dynamical properties by
considering the number of particles and the system length which is
defined as the position of the leftmost particle.  We use the same
formalism as in \cite{RefY,RefAY}, i.e.\ discrete time and parallel-update
scheme.  For generic hopping probability $p$, we will introduce a
domain wall prediction, checking it by Monte Carlo simulations.  In
the deterministic hopping case $p=1$, a rigorous analysis is
available, by using the generating function technique \cite{RefW}.

This paper is organized as follows.  In Sec.~\ref{model}, we define
the model as a discrete-time Markov process, and review its stationary
state based on \cite{RefAY}, which can be generalized to the
inhomogeneous injection case, see also App.~\ref{inhomo}.  In
Sec.~\ref{domain-wall}, we introduce a phenomenological argument on
how the number of particles $\langle N_t\rangle$ and the system length
$\langle L_t\rangle$ converge or diverge, showing
simulation results.  In Sec.~\ref{p=1}, we derive the asymptotic
behaviors of $\langle N_t\rangle$ and $\langle L_t\rangle$ rigorously
for $p=1$, imposing the initial condition that there is no particle on the
chain.  Section~\ref{conclusion} is devoted to the conclusion of this
paper.

%%%%%%%%%%%%%%%%%%%%%%%%%%%%%%%%%%%%
\section{Model} 
\label{model}

The EQP is defined on a semi-infinite chain where sites are labeled by 
natural numbers from right to left (Fig.~\ref{fig:eq}). 
Particles can enter the chain with probability $\alpha$ only at the left 
site next to the leftmost occupied site. A particle hops to its right nearest
neighbor site with probability $p$, if it is empty, and exits at the
right end of the chain with probability $\beta$.  If there is no
particle on the chain, a particle enters at site 1 with probability $\alpha$.
These transitions
occur simultaneously within one time step, i.e.\ we apply
the fully-parallel-update scheme.  

We formulate the EQP
as a discrete-time Markov process on the state
space
\begin{align}
S=\{\emptyset,1,10,11,100,101,110,111,1000,\dots\}
\end{align}
 where 0 and
1 correspond to unoccupied and occupied sites, respectively.  In
particular, $\emptyset$ denotes the state in which there is no
particle on the chain.  To simplify the notation we do not write the
infinite number of 0's located left to the leftmost 1.

Let us review the matrix product stationary state \cite{RefAY}
of the model,  which is a simple extension of that for systems
with a fixed system length \cite{RefBE}.
When 
\begin{align}\label{region}
\begin{cases} 
 \alpha\le\alpha_c=\frac{1-\sqrt{1-p}}{2} & \text{for } \beta>1-\sqrt{1-p},\\
 \alpha<\alpha_c=\frac{\beta(p-\beta)}{p-\beta^2} & \text{for } 
  \beta\le 1-\sqrt{1-p},
\end{cases}
\end{align}
the stationary state can be expressed as
\begin{align}\label{arrmpf1}
 P(\emptyset)=&\frac{1}{Z}\,, \\
\label{arrmpf2}
 P(1\tau_{L-1}\dots\tau_1)
 =&\frac{1}{Z}\left(\frac{\alpha}{p(1-\alpha)} \right)^L
   \langle W | DX_{\tau_{L-1}}\cdots X_{\tau_1} |V\rangle\,.
\end{align}
$X_1=D$ and $X_0=E$ are matrices, $\W $ is a row vector and $\V $ 
is a column vector satisfying the algebraic relations
\begin{align}\label{algrel}
\begin{split}
  EDEE =& (1-p)EDE + EEE + pEE, \\
  EDED =& EDD + EED + pED,\\
  DDEE =& (1-p)DDE + (1-p)DEE \\
               & \  + p(1-p)DE, \\
  DDED =& DDD + (1-p)DED + pDD, \\
 DDE\V  =& (1-\beta)DD\V  + (1-p)DE\V   \\
                & \  + p(1-\beta)D\V  ,  \\
 EDE\V  =& (1-\beta)ED\V  + EE\V  + pE\V , \\
 \W  DEE =&  (1-p) \W  DE, \\
 \W  DED =& \W  DD + p\W  D, \\
 DD\V  =& \frac{p(1-\beta)}{\beta}D\V ,  \ 
 ED\V   = \frac{p}{\beta}E\V , \\
 \W  EE =&  0, \ 
 \W  ED = p\W  D, \ 
\W  D \V  = \frac{p}{\beta}.
\end{split}
\end{align}
These relations are closely related to those for the stationary 
state of the parallel-update TASEP with ordinary open 
boundary condition \cite{RefERS}.
The normalization constant is expressed as
%\begin{align}
\begin{eqnarray}
\label{norcon}
%\begin{split}
 Z &=& 1+\sum_{L\ge 1} \left(\frac{\alpha}{p(1-\alpha)}\right)^L
  \W  D(D+E)^{L-1}\V  \\
&=& \frac{2(1-\alpha)\beta}{R-p+2(1-\alpha)\beta}
%\end{split}
%\end{align}
\end{eqnarray}
with $R=\sqrt{p(p-4\alpha(1-\alpha))}$.
The average number of particles $\langle N\rangle$
 and the average system length
 (the position of the leftmost particle) $\langle L\rangle$
 are calculated as
\begin{align}
\label{Nst}
 \langle N\rangle =&
    \frac{\alpha (1-\alpha)(p-2\alpha p+R)}{R(R-p+2(1-\alpha)\beta)}, \\
 \langle L\rangle =&
    \frac{\alpha p(R-p+2(1-\alpha))}{R(R-p+2(1-\alpha)\beta)}
\label{Lst}
\end{align}
in the stationary state.
Note that $\langle N\rangle $ and $\langle L\rangle $ diverge on the
critical line $ \alpha =\frac{1}{2}(1-\sqrt{1-p})$,
$\beta>1-\sqrt{1-p}$, where the stationary state exists.

A generalization of the model, where the entry probability depends
on the system length, also has a matrix product stationary
state, see App.~\ref{inhomo}.

%%%%%%%%%%%%%%%%%%%%%%%%%%
\begin{figure}[h]
\begin{center}
\centering
\includegraphics[height=62mm]{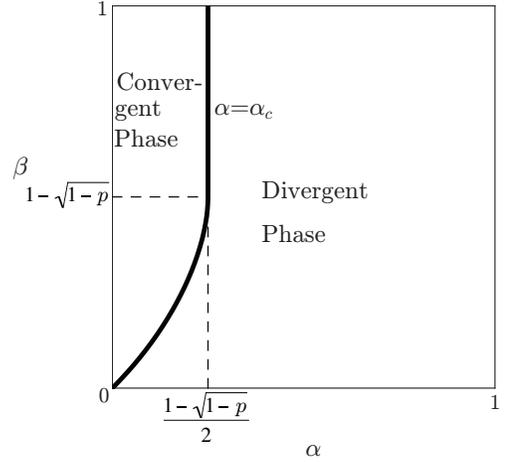}\\[0mm]
\caption{Phase diagram for the EQP.
The parameter space is divided into two regions
with and without the stationary state.}
\label{fig:pdgen}
\end{center}\end{figure}
%%%%%%%%%%%%%%%%%%%%%%%%%%

%%%%%%%%%%%%%%%%%%%%%%%%%%%%%%%%%%%%
\section{Domain wall picture and Monte Carlo simulations}\label{domain-wall}

In this section, we discuss the time evolution of the average number
 of particles $\langle N_t\rangle$ and the average system length
 $\langle L_t\rangle$ corresponding to the position of the leftmost
 particle.

In the ordinary open boundary case, where the length of the system is
fixed, a domain wall moves rightward or leftward, or exhibits a random
walk depending on the boundary parameters \cite{RefKSKS}.  In the same
way, we will discuss how the system length $\langle L_t\rangle$ moves.
We also observe how the average number of particles $\langle N_t\rangle$
changes as well.  The continuity equation
\begin{align}
  \label{rigoN}
   \langle N_{t+1}\rangle - \langle N_t\rangle
    &=J_t^{\rm in}-J_t^{\rm out}.
\end{align}
holds,
where $J_t^{\rm in}$ and $J_t^{\rm out}$
are the flows of particles entering and leaving the system, respectively.  
The inflow $J_t^{\rm in}$
is always $\alpha$, which is due to the fact that the site where
particles enter is by definition never blocked.  In other words, our
model is not a call-loss system.  Under the assumption that the
outflow $J^{\rm out}_t$ is independent of $t$, we have $\langle
N_t\rangle=(\alpha- J^{\rm out})t+\langle N_0\rangle$.  
In fact our simulations show that both $\langle N_t\rangle $ and
$\langle L_t\rangle$ decrease or increase linearly in time $t$
according to $\alpha<\alpha_c$ or $\alpha>\alpha_c$, respectively.

%%%%%%%%%%%%%%%%%%%%%%%%%%%%%%%%%%%%
\subsection{Convergent Phase}

\begin{figure}
\begin{center}
\centering
\includegraphics[width=1\columnwidth]{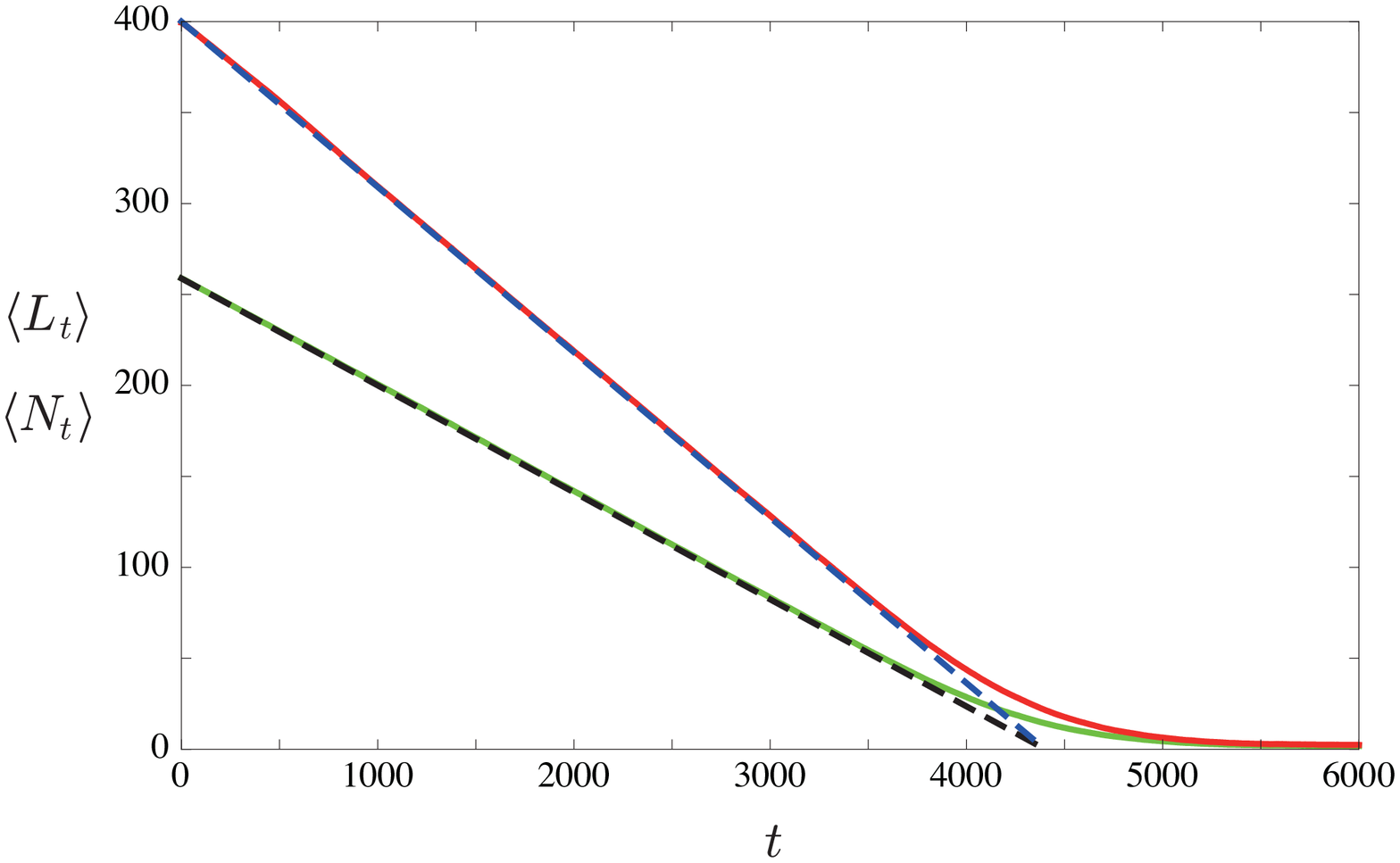}\\[3mm]
\includegraphics[width=1\columnwidth]{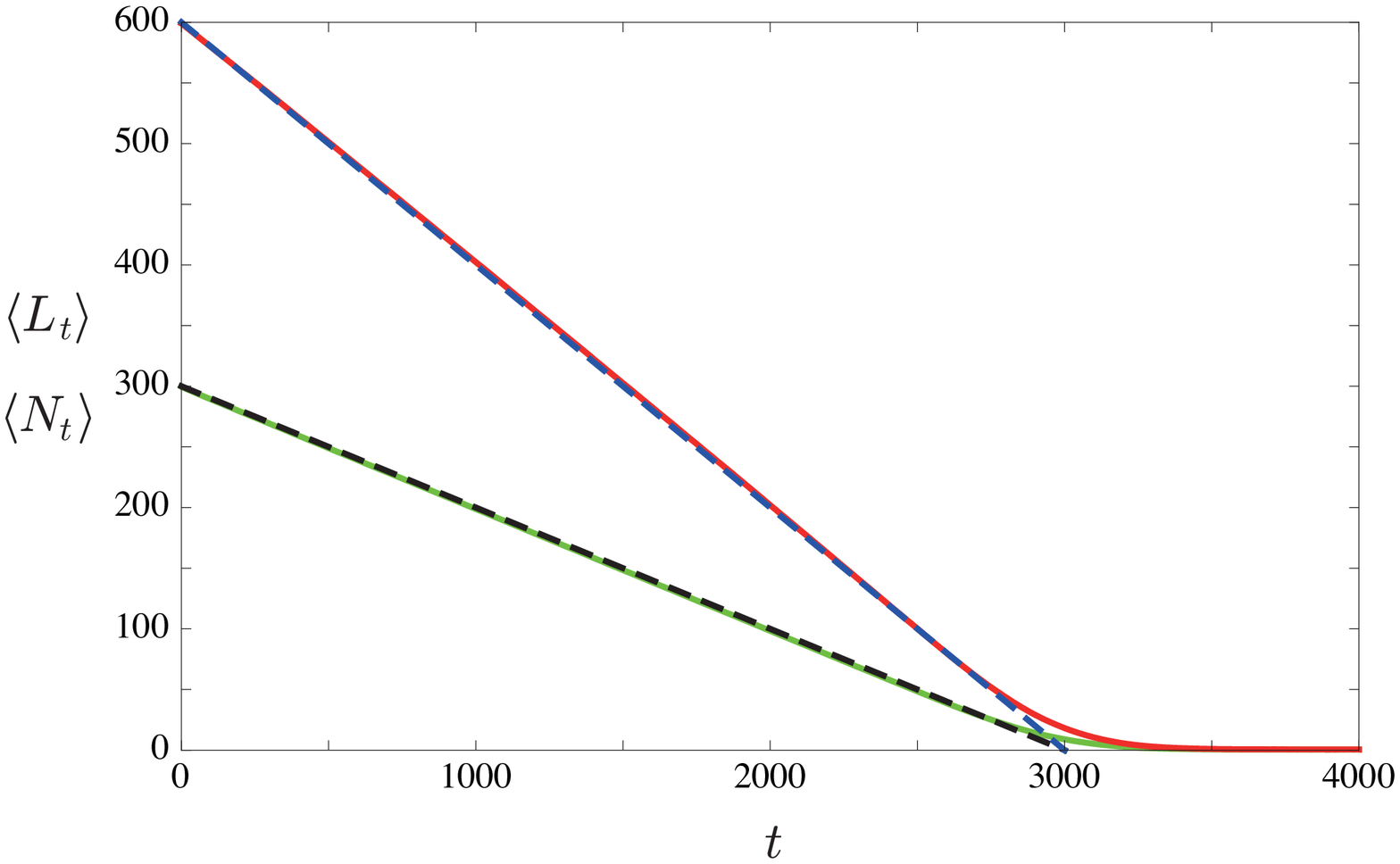}\\[3mm]
\caption{Dynamics in the HD-C (top) and MC-C (bottom) phases.
  The parameter values are chosen as $(\alpha,\beta,p)=(0.2,0.4,0.84)$
  and $(0.2,0.8,0.84)$, and the initial conditions as
  $(\langle N_0\rangle,\langle L_0\rangle) =(400\rho,400)$ and $(600\rho,600)$
  (with $\rho$ defined by \eqref{rho}), respectively.
  The green and red lines are data
  for $\langle N_t\rangle$ and $\langle L_t\rangle$
  obtained from Monte Carlo simulations, where 5000 samples are
  averaged. The black and blue lines correspond to the predictions of
  the domain wall theory.  Note that the asymptotic values are small
  but non-zero (see Eqs.~\eqref{Nst} and \eqref{Lst}): $\mathbf (\langle
  N_\infty\rangle, \langle L_\infty\rangle)=(1.70,2.28)$ and $\mathbf
  (0.42,0.56)$, respectively.  }
\label{fig:sim-conv} 
\end{center}
\end{figure}

When $\alpha<\alpha_c$, the system converges to the stationary state
\eqref{arrmpf1}, \eqref{arrmpf2}.  We impose the initial condition
that particles are distributed uniformly with density
\begin{align}\label{rho}
  \rho =&
  \begin{cases}
     \frac{1}{2}  &  \text{for } \beta>1-\sqrt{1-p}, \\
     \frac{p-\beta}{p-\beta^2}  & \text{for } \beta\le1-\sqrt{1-p}.
  \end{cases}
\end{align}
As in Fig.~\ref{fig:sim-conv},
  $\langle N_t\rangle $ and $\langle L_t\rangle $
  decrease linearly in time as
\begin{align}
\label{phenoN}
  \langle N_t\rangle \sim (\alpha-\Jout)t
     + \langle N_0\rangle , \\
  \langle L_t\rangle \sim \frac{\alpha-\Jout}{\rho}t
       + \langle L_0\rangle ,
\label{phenoL}
\end{align}
with
\begin{align}
\label{Jout}
\begin{split}
   J^{\rm out}
  &=\frac{1-\sqrt{1-4p\rho(1-\rho)} }{2} \\
  &=
   \begin{cases}
     \frac{1-\sqrt{1-p}}{2}  &  \text{for } \beta>1-\sqrt{1-p}, \\
     \frac{\beta(p-\beta)}{p-\beta^2}  &  \text{for } \beta\le 1-\sqrt{1-p}.
  \end{cases}
\end{split}
\end{align}
Since  $J^{\rm out}=\alpha_c$, we have $\alpha - J^{\rm out} <0$
which means that the domain wall moves rightward.
%%%%%%%%%%%%%%%%%%%%%%%%
%%%%%%%%%%%%%%%%%%%%%%%%

\begin{figure}
\begin{center}
\centering
\includegraphics[height=62mm]{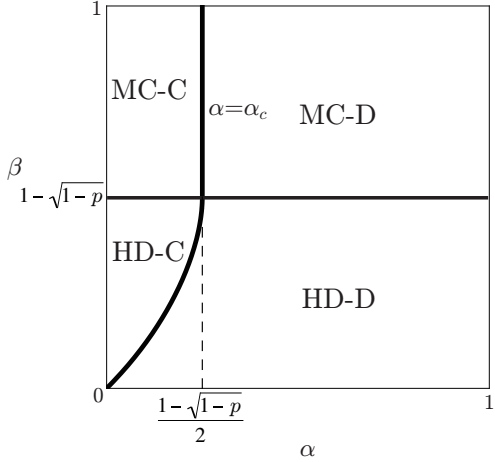}\\[0mm]
\caption{Subphases of the EQP.}
\label{fig:subphase} 
\end{center}
\end{figure}

According to the forms for
$J^{\rm out }$ and $\rho$, we call the phases
\begin{align}
& \text{MC-C:}\quad \alpha < \frac{1-\sqrt{1-p}}{2}
  \quad\text{and}\quad   \beta>1-\sqrt{1-p}, \\
& \text{HD-C:}\quad \alpha< \frac{\beta(p-\beta)}{p-\beta^2} 
  \quad\text{and}\quad \beta\le 1-\sqrt{1-p},
\end{align}
maximal-current-convergent (MC-C) and high-density-convergent (HD-C) phases,
respectively, see Fig.~\ref{fig:subphase}.
%%%%%%%%%%%%%%%%%%%%%%%%
%%%%%%%%%%%%%%%%%%%%%%%%

It should be noted that the outflow is given by Eqn.~\eqref{Jout} only
while $0\le t\lesssim \frac{\langle L_0\rangle}{\alpha-J^{\rm out}}$.
As $t\to\infty$, the outflow approaches $\alpha$, assuring that
$\langle N_t\rangle$ approaches the stationary value \eqref{Nst}.

\begin{figure}
\begin{center}
\centering
\includegraphics[width=1\columnwidth]{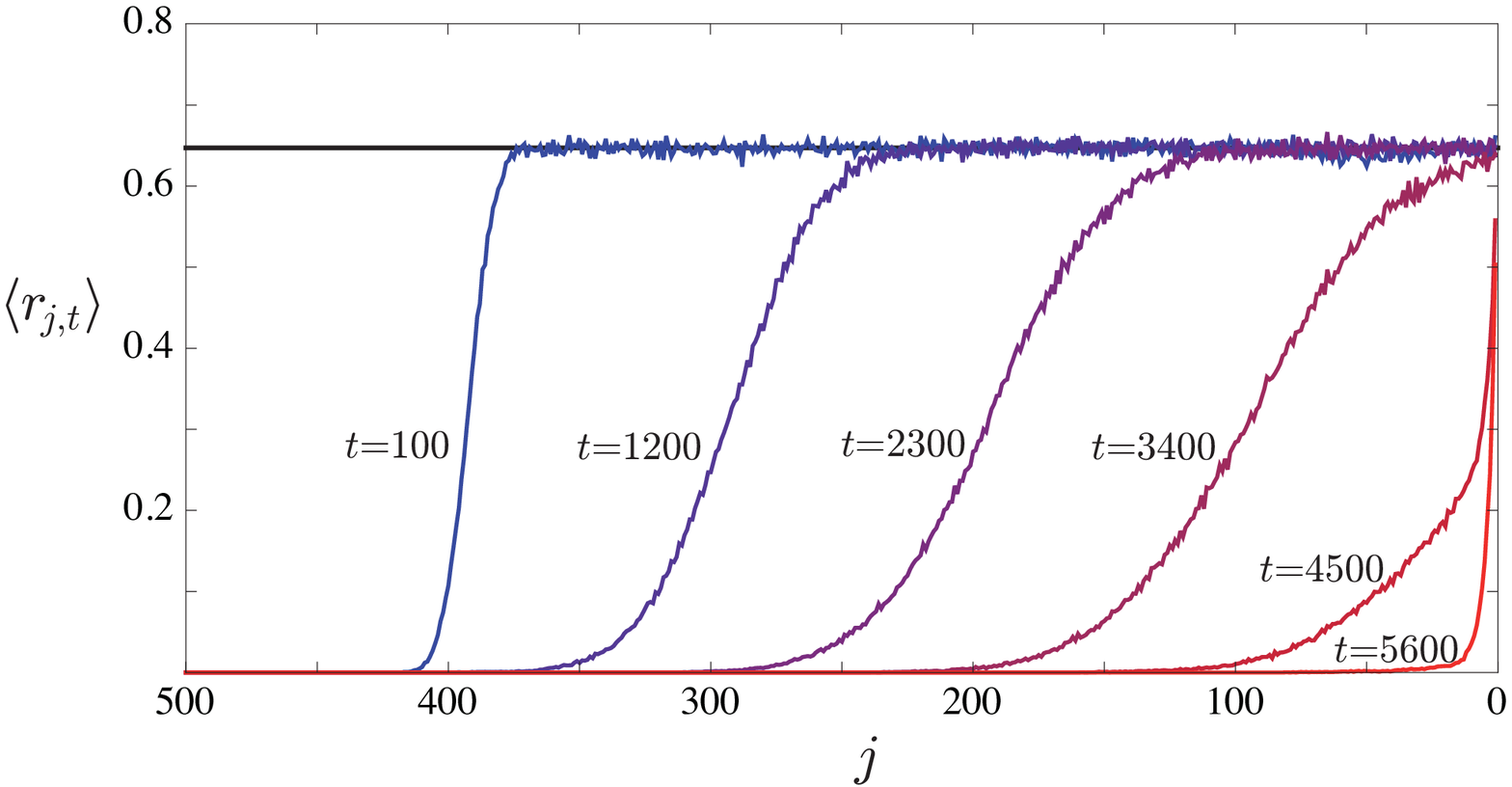}\\[3.2mm]
\includegraphics[width=1\columnwidth]{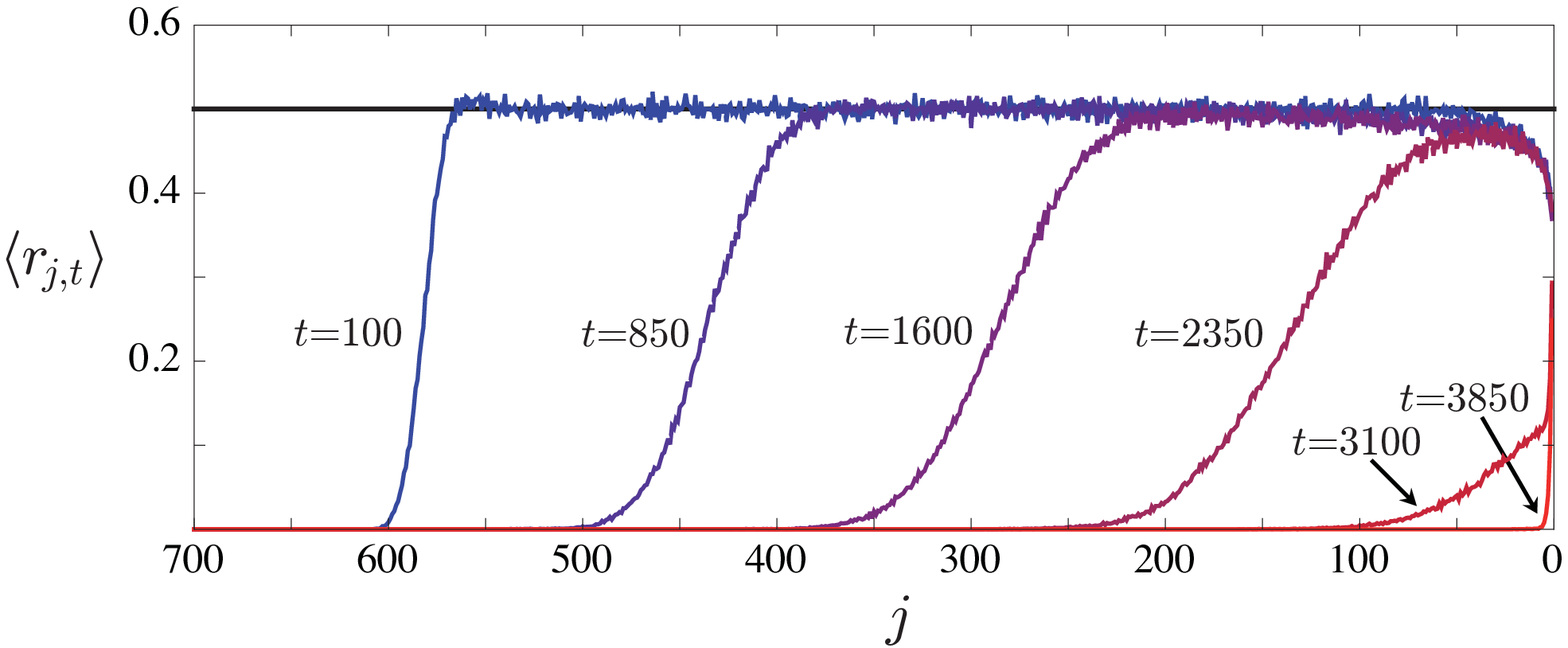}\\[2.8mm]
\caption{Density profiles ($\langle r_{j,t} \rangle$ of the $j$th 
  site at time $t$) in the HD-C (top) and MC-C (bottom) phases.  The
  parameters $(\alpha,\beta,p)$ and the initial conditions $(\langle
  N_0\rangle,\langle L_0\rangle)$ are set to the same values as in
  Fig.~\ref{fig:sim-conv}.  The colored snapshots are obtained by
  averaging 5000 samples of Monte Carlo simulations.
  The black lines represent the predicted densities $\rho$ according to
  Eqn.~\eqref{rho}.  }
\label{fig:sim-conv-rho} 
\end{center}
\end{figure}

Figure~\ref{fig:sim-conv-rho} shows density profiles in the HD-C and
MC-C phases.  We can observe that the bulk density keeps its initial
value \eqref{rho}.
%%%%%%%%%%%%%%%%%%%%%%%%

%%%%%%%%%%%%%%%%%%%%%%%%%%%%%%%%%%%%
\subsection{Divergent Phase}

%%%%%%%%%%%%%%%%%%%%%%%%
\begin{figure}\begin{center}
\includegraphics[width=1\columnwidth]{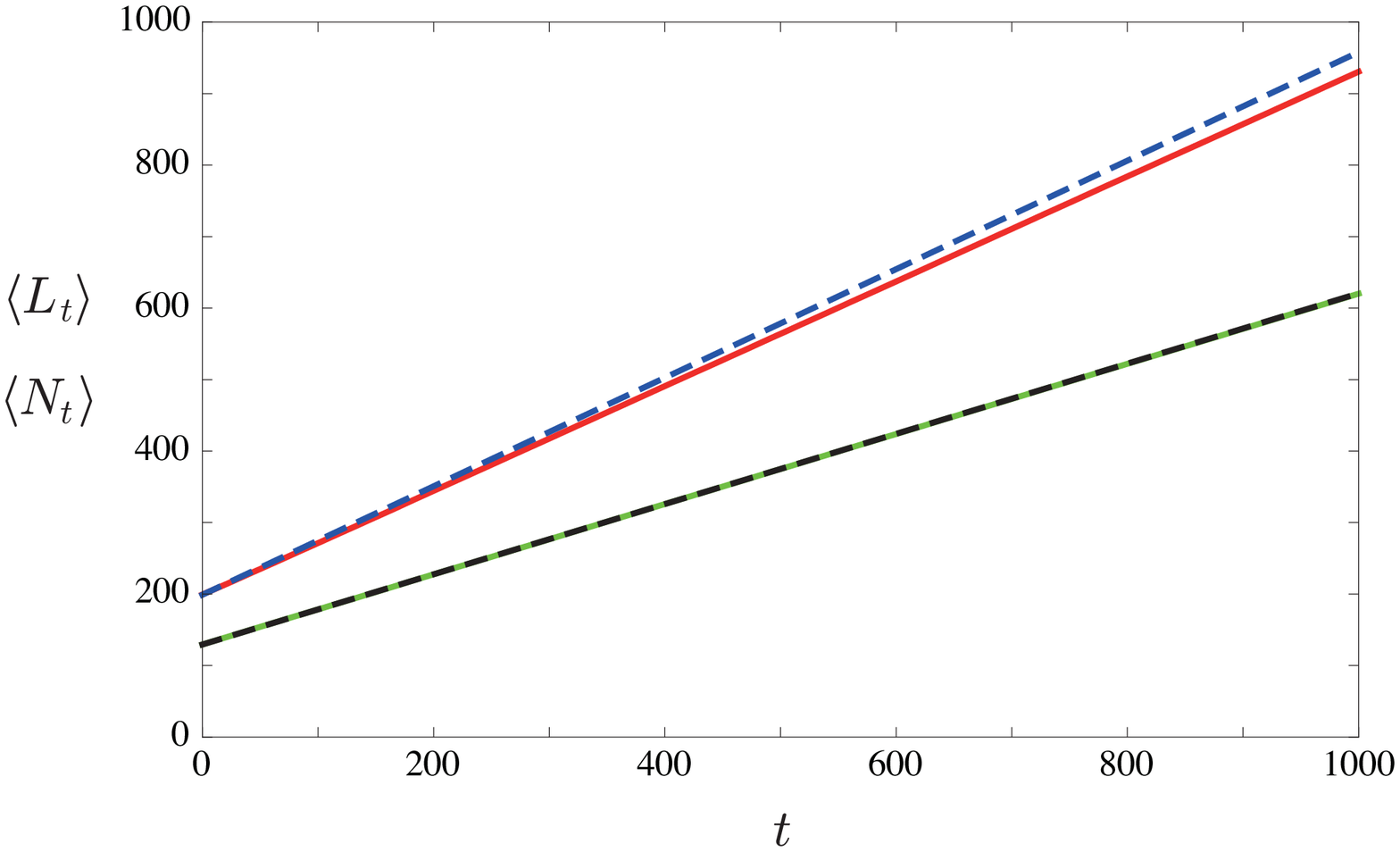}\\[3mm]
\includegraphics[width=1\columnwidth]{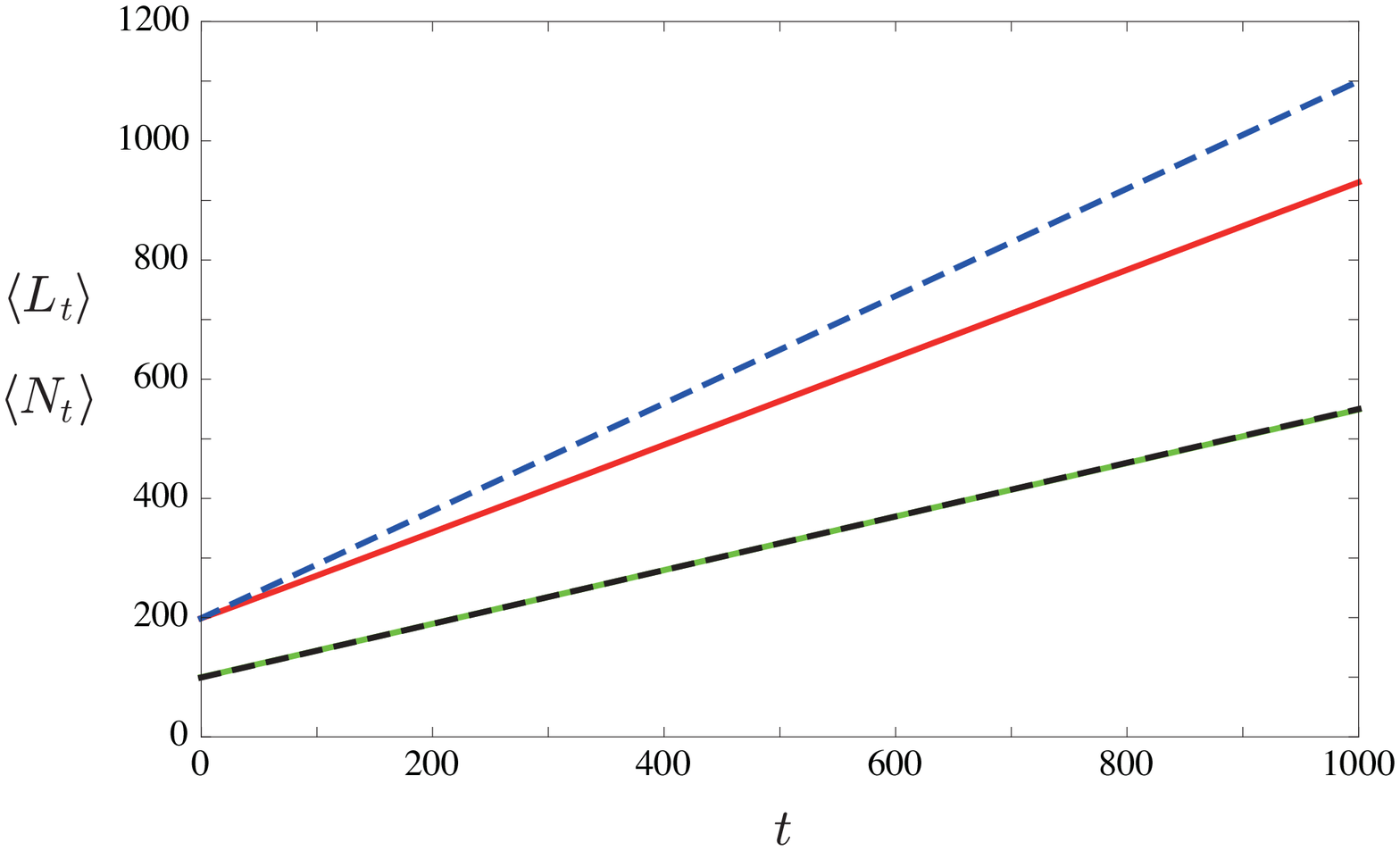}\\[3mm]
\caption{Dynamics in the HD-D (top) and MC-D (bottom) phases.
  The parameter values are chosen as $(\alpha,\beta,p)=(0.75,0.4,0.84)$
  and $(0.75,0.8,0.84)$, respectively, and the initial conditions as
  $(\langle N_0\rangle,\langle L_0\rangle) =(200\rho,200)$
  (with $\rho$ defined by \eqref{rho}).
  The green and red lines are data for
  $\langle N_t\rangle$ and $\langle L_t\rangle$
  obtained from Monte Carlo simulations, where 1000 samples are
  averaged.
  The black and blue lines correspond to the predictions of
  the domain wall theory.  }
\label{fig:sim-div} 
\end{center}\end{figure}
%%%%%%%%%%%%%%%%%%%%%%%%

When $\alpha>\alpha_c$, it is natural to expect that the domain wall
moves leftward, and the time evolutions of $\langle N_t\rangle$ and
$\langle L_t\rangle$ are expressed by Eqs.~\eqref{phenoN} and
\eqref{phenoL}, respectively, with the density \eqref{rho}
and the outflow \eqref{Jout}.
The simulations imply that this is true for $\langle N_t\rangle $ with
\begin{align}
\label{phenoNdiv}
  \langle N_t\rangle &\sim (\alpha-\Jout)t + \langle N_0\rangle , 
\end{align}
 but fails for $\langle L_t\rangle $, see Fig.~\ref{fig:sim-div}.
This failure is not unexpected since the predicted velocity 
$\frac{\alpha-\Jout}{\rho}$ can be greater than 1 whereas 
the length $\langle L_t\rangle $ can not be larger than
$t+\langle L_0\rangle$  by the definition of the model.
However, the simulation results (see Fig.~\ref{fig:sim-div}) indicate that 
\begin{align}
  \langle L_t\rangle &\sim Vt + \langle L_0\rangle ,
\label{phenoLdiv}
\end{align}
so that the prediction is qualitatively correct. 
The velocity $V$ satisfies
\begin{align}
  V &\to 0 \quad (\alpha\to\alpha_c), \\
  V &= 1 \quad (\alpha=1).
\end{align}
and  we have exactly $\langle L_t\rangle=t+\langle L_0\rangle$
when $\alpha=1$.
Moreover, when $p=1$, we will show in the next section
that
\begin{align}
  V=\alpha-\beta+\alpha\beta
   =\frac{ \alpha -\frac{\beta}{1+\beta}}{\frac{1}{1+\beta}}
   =\frac{ \alpha - J^{\rm out}}{\rho}.
\end{align}

Equations \eqref{phenoNdiv} and \eqref{phenoLdiv} can be regarded as
the asymptotic behaviors
\begin{align}
\label{phenoNasym}
  \langle N_t\rangle &= (\alpha-\Jout)t + o(t), \\
  \langle L_t\rangle &= V t  + o(t).
\label{phenoLasym}
\end{align}

In the same way as in the convergent phase,
we call the subphases
\begin{align}
&\text{MC-D:}\quad \alpha> \frac{1-\sqrt{1-p}}{2} 
 \quad  \text{and} \quad \beta>1-\sqrt{1-p}, \\
&\text{HD-D:}\quad \alpha> \frac{\beta(p-\beta)}{p-\beta^2} 
 \qquad \text{and} \qquad \beta\le1-\sqrt{1-p},
\end{align}
maximal-current-divergent (MC-D) and high-density-divergent (HD-D) phases,
respectively.
Note that the densities in the MC-D and HD-D phases
 are higher than (or equal to) $\frac{1}{2}$ and $\frac{p-\beta}{p-\beta^2}$,
 respectively.

It is difficult to predict how $\langle N_t\rangle$
or $\langle L_t\rangle$ behaves just on the critical line $\alpha=\alpha_c$.
For $p=1$, however, we will find in the next section
diffusive behavior on the critical line as
%\begin{align}
\begin{eqnarray}
  \label{phenoN_critical}
  \langle N_t\rangle &=& D_N \sqrt{t} + o(\sqrt{t}), \\
  \langle L_t\rangle &=& D_L \sqrt{t} + o(\sqrt{t})
%\end{align}
\end{eqnarray}
with constants $D_N$ and $D_L$.

%%%%%%%%%%%%%%%%%%%%%%%%%%%%%%%%%%%%
\section{Asymptotic behaviors for $p=1$}\label{p=1}

%%%%%%%%%%%%%%%%%%%%%%%%%%
\begin{figure}
\begin{center}
\includegraphics[width=0.7\columnwidth]{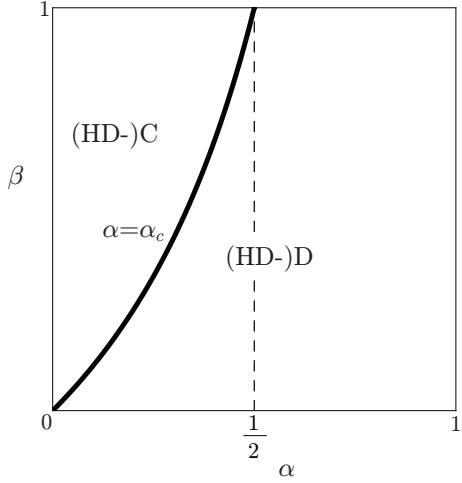}\\[0mm]
\end{center}
\caption{Phase diagram for $p=1$.}
\label{fig:pd184}
\end{figure}
%%%%%%%%%%%%%%%%%%%%%%%%%%

In this section we investigate the asymptotic behaviors of $\langle
N_t\rangle$ and $\langle L_t\rangle$ rigorously for $p=1$.  Thanks to
the deterministic particle hopping we can obtain the generating
functions of $\langle N_t\rangle$ and $\langle L_t\rangle$.  For
simplicity, we impose the initial condition $\emptyset$
(there is no particle in the system at time $t=0$),
and reset the state space as
\begin{align}
\begin{split}
 \widetilde{S} 
  &:=\{\emptyset,1,10,11,101,110,111,1010,1011,\dots\} \\
  &=\left\{\tau\in S|\tau\ \text{does not contain sequence }
     00   \right\}.
\end{split}
\end{align}
Note that, for $p=1$, the sequence 00 never appears if the system
starts from the initial condition $\emptyset$.  In this case, the MC-D
and MC-C phases vanish from the phase diagram, see
Fig.~\ref{fig:pd184}.

We first consider the number of particles, borrowing the
classification from \cite{RefY} as
\begin{align}
\begin{split}
  P_t^A(N) =&
  {\rm Prob}\left[\begin{array}{l}
     \text{\# of particles is $N$ at time $t$} \\
  \wedge \text{ site 1 is occupied at time $t$}
   \end{array} \right],
\end{split}
\\
\begin{split}
  P_t^B(N) =&
  {\rm Prob}\left[\begin{array}{l}
     \text{\# of particles is $N$ at time $t$} \\
  \wedge \text{ site 1 is empty at time $t$}
   \end{array} \right]
\end{split}
\end{align}
for $N\in\mathbb Z_{\ge 0}$ with $P_t^A(0)\equiv 0$.  These
probabilities are governed by the following master equation, which was
found in \cite{RefY}:

\begin{align}\label{master_for_n}
\begin{split}
  P_{t+1}^A(1)=&
    (1-\alpha)(1-\beta)P_t^A (1) + \alpha P_t^B(0) \\
     &\quad\quad\quad\quad\quad+ (1-\alpha)P_t^B(1) ,
\end{split}
\\
\begin{split}
  P_{t+1}^A(N)=&
    \alpha(1-\beta)P_t^A(N-1) + \alpha P_t^B(N-1) \\
     &\!\!\!\!\!\!\!\!\!+ (1-\alpha)(1-\beta)P_t^A(N) + (1-\alpha) P_t^B(N),
\end{split}
\\
  P_{t+1}^B(0)=&
    (1-\alpha) P_t^B(0) + (1-\alpha)\beta P_t^A(1), \\
  P_{t+1}^B(N)=&
    \alpha\beta P_t^A(N) + (1-\alpha)\beta P_t^A(N+1) .
\end{align}
This simplification is due to the deterministic hopping $p=1$.
We choose the initial condition such that
\begin{align}\label{initial}
   P^B_0(0)=1,\quad
   P^A_0(N)=P^B_0(N)=0\quad (N\in\mathbb N).
\end{align}

We will check that the average number of particles $\langle
N_t\rangle$ converges to the stationary value \eqref{Nst} when
$\alpha<\frac{\beta}{1+\beta}$, and show that $\langle N_t\rangle$
behaves as Eqn.~\eqref{phenoNasym} when
$\alpha>\frac{\beta}{1+\beta}$.  We will also show that $\langle
N_t\rangle$ exhibits diffusive behavior on the critical line
$\alpha=\frac{\beta}{1+\beta}$.

We define the generating functions of $P^A_t(N)$ and $P^B_t(N)$ as
%\begin{align}
\begin{eqnarray}
   G^A_z(N) &=& \sum_{t\ge 0} P_t^A(N) z^t,\\
   G^B_z(N) &=& \sum_{t\ge 0} P_t^B(N) z^t,
%\end{align}
\end{eqnarray}
for $|z|<1$. 
Noting the initial condition \eqref{initial}, we find
\begin{align}
\begin{split}
\label{GA1}
   &G_z^A(1)=
    (1-\alpha)(1-\beta)zG_z^A (1) + \alpha zG_z^B(0) \\
     &\quad\quad\quad\quad\quad+ (1-\alpha)zG_z^B(1) ,
\end{split}
\\
\begin{split}
\label{GAN}
   &G_z^A(N)=
    \alpha(1-\beta)zG_z^A(N-1) + \alpha zG_z^B(N-1) \\
     &\quad+  (1-\alpha)(1-\beta)zG_z^A(N) + (1-\alpha)zG_z^B(N),
\end{split}
\\
\label{GB0}
   & G_z^B(0) -1 =
    (1-\alpha) zG_z^B(0) + (1-\alpha)\beta zG_z^A(1), \\
\label{GBN}
  &G_z^B(N)=
    \alpha\beta zG_z^A(N) + (1-\alpha)\beta zG_z^A(N+1) .
\end{align}
From Eqn.~\eqref{GB0}, we have
\begin{align}
 G_z^B(0) &= \frac{ 1+(1-\alpha)\beta z G_z^A(1) }{ 1-(1-\alpha)z }.
\end{align}
Inserting this and Eqn.~\eqref{GBN} into Eqs.~\eqref{GA1} and
\eqref{GAN}, we get a recurrence formula for $G_z^A(N)$ as
\begin{widetext}
\begin{align}\label{GA2=}
  &G_z^A(2)
  = - \frac{\alpha}{ (1-\alpha)^2\beta (1-(1-\alpha)z) z } 
    +    \frac{1 - (1-\alpha)(2-\beta)z
              + (1-\alpha)(1-\alpha-\beta-\alpha\beta)z^2
              + (1-\alpha)^2 \alpha\beta z^3}
    { (1-\alpha)^2\beta (1-(1-\alpha)z) z^2 }
    G_z^A(1) \\
\begin{split}\label{recursion}
  &G_z^A(N+1)
  = - \frac{1 -  (1-\alpha)(1-\beta)z - 2(1-\alpha)\alpha\beta z^2}
  { (1-\alpha)^2 \beta z^2 } G_z^A(N)
   + \frac{ \alpha (1-\beta) + \alpha^2\beta z}
  { (1-\alpha)^2 \beta z } G_z^A(N-1) \\
  &\quad \quad \quad \quad \ 
   \Big( =: x G_z^A(N) + y G_z^A(N-1) \Big)
     \quad\quad(N\in\mathbb Z_{\ge 2}).
\end{split}
\end{align}
\end{widetext}
The recurrence formula~\eqref{recursion} has the following solution:
\begin{align}
\begin{split}
   G_z^A(N) =& \frac{\lambda_+^{N-1}}{\lambda_+ - \lambda_-}
   \left( G_z^A(2) - \lambda_- G_z^A(1) \right)  \\
  & - \frac{\lambda_-^{N-1}}{\lambda_+ - \lambda_-}
   \left( G_z^A(2) - \lambda_+ G_z^A(1) \right) , 
\end{split}
\end{align}
where
\begin{equation}
\lambda_\pm =
\frac{1-(1-\alpha ) (1-\beta )z - 2(1-\alpha ) \alpha\beta z^2 \pm r}
{2 (1-\alpha )^2 \beta z^2 } 
\end{equation}
with
\begin{equation}
r =\sqrt{ (1-(1-\alpha)(1-\beta)z)^2 - 4 (1-\alpha)\alpha\beta z^2 }
\end{equation}
are solutions to $ \lambda^2=x\lambda+y$ 
with $x$ and $y$ as defined in Eqn.~\eqref{recursion}.
Restricting the ``initial condition'' $G^A_z(1)$ and $G^A_z(2)$ such that
\begin{align}
   \left|G^A_z (N)\right| < \sum_{t\ge0}  |z|^t  =\frac{1}{1-|z|} ,
\end{align}
we have
\begin{align}
 G_z^A(2) - \lambda_- G_z^A(1)  = 0.
\end{align}
(Note that $0<|\lambda_-|<1<|\lambda_+|$.)  From this constraint and the
relation \eqref{GA2=} $G^A_Z(1)$ is determined, and we find
\begin{widetext}
\begin{align}
G_z^A(N) &= \lambda_-^{N-1} G_z^A(1) =
\lambda_-^{N-1} \frac{-1+(1-\alpha)(2-\beta)z 
-(1-\alpha)(1-\alpha-\beta-\alpha\beta)z^2+(1-(1-\alpha)z)r}
{2(1-\alpha)^2\beta^2(1-z)z^2} \quad (N\in \mathbb N), \\
G_z^B(0) &= \frac{-1+r+(1-\alpha)(1+\beta)z}{2(1-\alpha)\beta z(1-z)},
\qquad
G_z^B(N)  = \left(\alpha+(1-\alpha)\lambda_-\right)\beta z
\lambda_-^{N-1} G_z^A(1) \quad (N\in \mathbb N).
\end{align}
Then we obtain the generating function $G_z(N)$
of the probability that the number of particles is $N$ as
\begin{align}
G_z(0) =&~G_z^B(0),\quad
G_z(N) = G_z^A(N) + G_z^B(N) = g(z)\lambda_-^N, \\
g(z)=&\,
\frac{\beta-1
+\left(1-\alpha-\beta+\beta^2-\alpha\beta^2\right)z
+(1-\alpha)(1-\beta)\beta z^2  + r (1-(1-z)\beta)}
{2\beta(1-\alpha)(1-(1-z\alpha)\beta)(1-z)z}
\quad (N\in \mathbb N).
\end{align}
\end{widetext}
We also introduce the generating function ${\mathcal G}_{z\zeta}$ of
the generating function $G_z(N)$ as
\begin{align}
\begin{split}
{\mathcal G}_{z\zeta}
&= \sum_{N\ge 0}G_N(z)\zeta^N  
 = G_0(z) + \sum_{N\ge 1}g(z)(\lambda_-\zeta)^N \\
&= G_0(z) + g(z)\frac{\lambda_-\zeta}{1-\lambda_-\zeta},
\end{split}
\end{align}
and $K_z$ of the average number of particles as
\begin{align}
\begin{split}
  K_z
  &= \sum_{t\ge0} z^t \langle N_t\rangle 
   = \sum_{t\ge0} z^t \sum_{N\ge 0}N
     \left(P^A_t(N)+ P^B_t(N) \right)  \\
  &= \left.\frac{\partial}{\partial\zeta}
     {\mathcal G}_{z\zeta}\right|_{\zeta=1}
   = \frac{z(2\alpha-1+z(1-\alpha)(1-\beta)+r)}{2(1-z)^2(1+z\beta)}.
\end{split}
\end{align}

The asymptotic behavior of $\langle N_t\rangle$ is determined by
the degree of the singularity $z=1$ of $K_z$ \cite{RefW}. 
When $\alpha<\frac{\beta}{1+\beta}$, we find
\begin{align}
  (1-z)K_z\Big|_{z\to 1} =\frac{\alpha(1-\alpha)}{\beta-\alpha-\alpha\beta},
\end{align}
and $\langle N_t \rangle $ converges as
\begin{align}
  \langle N_t \rangle \to
   \frac{\alpha(1-\alpha)}{\beta-\alpha-\alpha\beta}
   \quad\quad (t\to\infty).
\end{align}
Of course this limit value agrees with the stationary value \eqref{Nst}
with $p=1$.
When $\alpha>\frac{\beta}{1+\beta}$, we find
\begin{align}
  (1-z)^2K_z\Big|_{z\to 1} = \alpha -\frac{\beta}{1+\beta} ,
\end{align}
and $\langle N_t \rangle $ behaves as
\begin{align}
  \langle N_t \rangle &=
  \left( \alpha -\frac{\beta}{1+\beta} \right) t +o(t) \\
  &=  (\alpha-J^{\rm out}) t +o(t)     \quad\quad  (t\to\infty).
\end{align}
When $\alpha=\frac{\beta}{1+\beta}$, we find
\begin{align}
  (1-z)^{\frac{3}{2}}K_z\Big|_{z\to 1}
 = \sqrt{ \frac{\beta}{(1+\beta)^3} } ,
\end{align}
and $\langle N_t \rangle $ behaves as
\begin{align}
  \langle N_t \rangle =
  2\sqrt{ \frac{\beta t}{\pi(1+\beta)^3} }
      +o\left(\sqrt{t}\right)
   \quad \quad (t\to\infty).
\end{align}

Now we turn to the behavior of the length of the system
 (the position of the leftmost particle).
Let $Q_t(L)$ be the probability that the system length
is $L$ at time $t$.  The probability $Q_t(L)$ is governed by
\begin{eqnarray}
\label{Q0}
  Q_{t+1}(0) &=& (1-\alpha)Q_t(0) + \beta (1-\alpha)Q_t(1), \\
  Q_{t+1}(L) &=&  \alpha Q_t(L-1) + (1-\alpha) (1-\beta)Q_t(L) \nonumber\\
           &&+  (1-\alpha) \beta Q_t(L+1)
\label{QL}
\end{eqnarray}
for $L\in\mathbb N$.  The first equation means that, if there is no
particle at time $t+1$, there is no particle at time $t$ and no
particle enters (with probability $1-\alpha$), or there is only one
particle on the rightmost site at time $t$ which leaves the system and
no particle enters (with probability $(1-\alpha)\beta$).  The second
equation is derived in App.~\ref{derive-QL}.

In the same way as for the number of particles, we define the
generating function $M_z(L)=\sum_{t\ge 0} z^tQ_t(L)\ (|z|<1)$.  Noting
the initial condition $ Q_0(0)=1$ and $Q_0(L)=0\ (L\in \mathbb N)$, we find
\begin{align}
\label{M0}
\begin{split}
  M_z(1) =&\, \frac{1-(1-\alpha)z}{(1-\alpha)\beta z} M_z(0)
           -\frac{1}{(1-\alpha)\beta z},
\end{split}
\\
\begin{split}
 M_z(L+1)
 =&\, \frac{1-(1-\alpha)(1-\beta)z}{(1-\alpha)\beta z}M_z(L) \\
  &\,   -\frac{\alpha}{(1-\alpha)\beta} M_z(L-1)  \\
 \big( =:&\, X M_z(L) + Y M_z(L-1) \big) .
\end{split}
\label{ML}
\end{align}
The solution to the recurrence formula \eqref{ML} is
\begin{align}
  \begin{split}
   M_z(L) =& \frac{\Lambda_+^L }{\Lambda_+ - \Lambda_-}
   \left( M_z(1) - \Lambda_- M_z(0) \right)  \\
  & - \frac{\Lambda_-^L }{\Lambda_+ - \Lambda_-}
   \left( M_z(1) - \Lambda_+ M_z(0) \right),
\end{split}
\end{align}
where $ \Lambda_\pm =
 \frac{1-(1-\alpha)(1-\beta)z\pm r}{2(1-\alpha)\beta z}$
are the solutions to
$  \Lambda^2 = X \Lambda +Y $ 
with $X$ and $Y$ as defined in Eqn.~\eqref{ML}.
Due to the condition $\left|M_z(L)\right|<\frac{1}{1-|z|}$,
the ``initial condition'' must be restricted as
\begin{align}
  M_z(1) - \Lambda_- M_z(0) = 0.
\end{align}
(Note that $0<|\Lambda_-|<1<|\Lambda_+|$.)
Thus we find
\begin{align}
  M_z(L) = M_z(0) \Lambda_-^L
    = \frac{2}{1-(1-\alpha)(1+\beta)z+r} \Lambda_-^L.
\end{align}
The generating function of the generating function is 
calculated as
\begin{align}
  {\mathcal M}_{z\xi}
  =\sum_{L\ge0} \xi^L M_z(L) = \frac{M_z(0)}{1-\xi\Lambda_-},
\end{align}
and that of the average system length as
\begin{align}
\begin{split}
 S_z &= \sum_{t\ge 0} z^t \langle L_t\rangle
  =\sum_{t\ge 0} z^t \sum_{L\ge 0} L Q_t(L)  \\
&= \left.\frac{\partial}{\partial\xi}
  {\mathcal M}_{z\xi}\right|_{\xi=1}
= \frac{-1 +(1+\alpha-\beta+\alpha\beta)z+r}
{2(1-z)^2}.
\end{split}
\end{align}

The asymptotic behavior of $\langle L_t\rangle$ 
is determined by the degree of the singularity $z=1$ of $S_z$.
When $\alpha<\frac{\beta}{1+\beta}$, we find
\begin{align}
  (1-z)S_z\Big|_{z\to 1} =
\frac{\alpha}{\beta-\alpha-\alpha\beta},
\end{align}
and $\langle L_t \rangle $ converges as
\begin{align}
  \langle L_t \rangle \to
   \frac{\alpha}{\beta-\alpha-\alpha\beta}
   \quad\quad (t\to\infty).
\end{align}
Again this limit value agrees with the stationary value
\eqref{Lst} with $p=1$. When $\alpha>\frac{\beta}{1+\beta}$, we find
\begin{align}
  (1-z)^2S_z\Big|_{z\to 1} = \alpha - \beta +\alpha \beta,
\end{align}
and $\langle L_t \rangle $ behaves as
\begin{align}
  \langle L_t \rangle &=
  \frac{ \alpha -\frac{\beta}{1+\beta}}
    {\frac{1}{1+\beta}} t   + o(t)   \\
  &=  \frac{\alpha-J^{\rm out}}{\rho}t + o(t)  \quad\quad  (t\to\infty ).
\end{align}
When $\alpha=\frac{\beta}{1+\beta}$, we find
\begin{align}
  (1-z)^{\frac{3}{2}}S_z\Big|_{z\to 1}
 = \sqrt{ \frac{\beta}{1+\beta} } ,
\end{align}
and $\langle L_t \rangle $ behaves as
\begin{align}
  \langle L_t \rangle =
  2\sqrt{ \frac{\beta t}{\pi(1+\beta)} }
      +o\left(\sqrt{t}\right)
   \quad \quad (t\to\infty).
\end{align}

%%%%%%%%%%%%%%%%%%%%%%%%%%%%%%%%%%%%
\section{Conclusion}\label{conclusion}

We have investigated the dynamical properties of
the EQP, a queueing process with excluded-volume effect.
The model can be interpreted as a TASEP
with varying length. Using generating function techniques and a
phenomenological domain wall theory we have derived analytical
predictions for the time-dependence of the number of particles
$\langle N_t\rangle$ and the average system length $\langle
L_t\rangle$.

We found that the two phases observed previously can be divided in
subphases. The convergent phase, where the system length remains
finite, consists of high-density and maximal current subphases. The
same is true for the diverging phase, where the system length becomes
infinite in the long time limit.

By comparing with Monte Carlo simulations it was found that the
predications of the domain wall theory for the dynamical behavior are
at least qualitatively correct, i.e.\ $\langle L_t\rangle$ and
$\langle N_t\rangle$ converge or diverge linearly in time.  Moreover
they are in good agreement even quantitively in the convergent phase.
In the divergent phase, the predicted velocity for $\langle
N_t\rangle$ appears to be correct whereas deviations from the domain
wall theory can be observed for $\langle L_t\rangle$.

For $p=1$, we derived exact analytical results for the behaviors of
$\langle N_t\rangle$ and $\langle L_t\rangle$ by using the generating
function method.  We showed the linearity of their dynamics in the
divergent phase as predicted by the domain wall theory.  We found 
diffusive behavior on the critical line as well.

The simple approach presented here does not provide a good
expression for the velocity $V$ of the system length
($\langle L_t\rangle\sim Vt$) in the divergent phase.
Here further analysis of the detailed density profiles in the divergent 
phase may be helpful. A first step would be the numerical determination
of $V$ to get a better understanding of its dependence on the 
parameters $\alpha,\beta$ and $p$.
Another way to settle the problem is extending the exact result to
$p<1$ case, which may be difficult but very worthwhile.

%%%%%%%%%%%%%%%%%%%%%%%%%%%%%%%%%%%%
\appendix

\section{Stationary State for Inhomogeneous Injection Case\label{inhomo}}

Here we consider the stationary state for a generalized model
where the entry probability depends on the system length.
A new particle enters the system with probability $\alpha_L$
if the leftmost occupied site is $L$, or $\alpha_0$ if there is no
particle on the chain.  The stationary state of this generalized model
can be written in the following
 matrix product form with the same matrices and
vectors ($D,E,\langle W|$ and $|V\rangle$):
\begin{align}
&  P(\tau_L\cdots\tau_1)  =
  \frac{1}{Z}
  \frac{\prod_{j=0}^{L-1}\alpha_j}{ p^L\prod_{j=1}^{L} (1-\alpha_j)}
   \W  X_{\tau_L} \cdots X_{\tau_1} \V ,  \\
& Z= 1 + \sum_{L\ge 1} 
 \frac{\prod_{j=0}^{L-1}\alpha_j}{ p^L\prod_{j=1}^{L} (1-\alpha_j)}
\W  D (D+E)^{L-1}  \V  .
\end{align}
This can be proved in the same way as in the homogeneous case
$\alpha_L=\alpha$, see \cite{RefAY}.

%%%%%%%%%%%%%%%%%%%%%%%%%%%%%%%%%%%%
\section{Derivation of Eqn.~\eqref{QL}}\label{derive-QL}

To derive  Eqn.~\eqref{QL}, we show
\begin{align}
\begin{split}
\label{betaQ}
 &\beta Q_t(L) \\
 &=
  {\rm Prob}\left[\begin{array}{l}
     \text{the system length is $L$ at time $t$} \\
  \wedge \text{ ($L-1$)th site is empty at time $t$}
   \end{array} \right],
\end{split}
\\
\begin{split}
\label{1-betaQ}
 & (1-\beta) Q_t(L) \\
 & =
  {\rm Prob}\left[\begin{array}{l}
     \text{the system length is $L$ at time $t$} \\
  \wedge \text{ ($L-1$)th site is occupied at time $t$}
   \end{array} \right],
\end{split}
\end{align}
for $L\in\mathbb N_{\ge 2}$

\begin{figure}\begin{center}
\centering
\includegraphics[width=1\columnwidth]{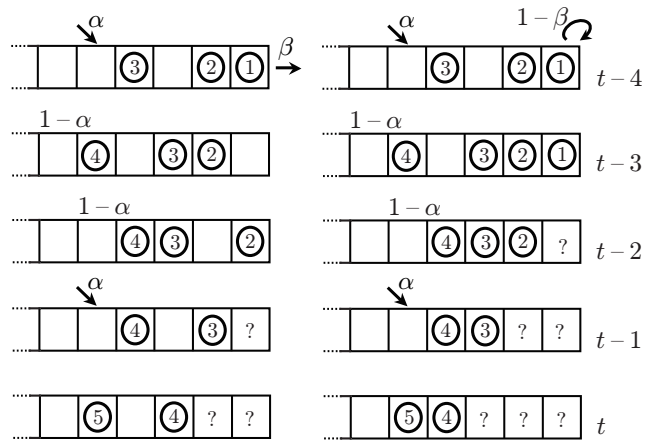}
\caption{Example for transitions from $1011$ at time $t-4$ 
  to the states in which the system length is 5 and the 4th site is
  empty (left) or occupied (right) at time $t$.}
\label{Transition-Example}
\end{center}\end{figure}

In the deterministic hopping case $p=1$
with the initial condition $\emptyset$,
the sequence 00 (except the infinite number of 0's
left to the leftmost particle) never appears,
and holes necessarily ``hop'' leftward.
Thus the first site must be occupied by a particle at time $t-L+1$ if
the system length at time $t$ is $L$.  The local state (empty or
occupied) of the $(L-1)$th site at time $t$ depends only on whether the
first particle exits or not at time $t-L+1$.  Let the number of
particles be $n$ at time $t-L+1$, and label the particles by natural
numbers as in Fig.~\ref{Transition-Example}.  We can show by induction
that the $\ell$th particle is on the $\ell$th site at time $t-L+\ell$.
Thus we find that $L-n$ new particles should enter the system during
$[t-L+1,t-1]$.

We obtain (the summation of) the transition probability from a
configuration $S^{(t-L+1)}$ at time $t-L+1$ to the states in which the
system length is $L$ and
the $(L-1)$th site is empty or occupied at time
$t$ is given by
\begin{align}
\begin{split}
&\sum_{S^{(t)},\dots,S^{(t-L+2)}\in \widetilde{S}
 \atop |S^{(t)}|=L, (S^{(t)})_{L-1}=0}
\prod_{t'=t-L+1}^{t-1} W(S^{(t')}\to S^{(t'+1)}) \\
& \ =\left(\begin{array}{c} L-1 \\ L-n \end{array}\right)
\alpha^{L-n}(1-\alpha)^{n-1}\beta,
\end{split}
\\
\begin{split}
&\sum_{S^{(t)},\dots,S^{(t-L+2)}\in \widetilde{S}
 \atop |S^{(t)}|=L,(S^{(t)})_{L-1}=1}
\prod_{t'=t-L+1}^{t-1} W(S^{(t')}\to S^{(t'+1)}) \\
&\ =\left(\begin{array}{c} L-1 \\ L-n \end{array}\right)
\alpha^{L-n}(1-\alpha)^{n-1}(1-\beta),
\end{split}
\end{align}
respectively, where $W(S^{(t')}\to S^{(t'+1)})$, $|S^{(t')}|$ 
and $(S^{(t')})_j$ denote the transition probability from 
$S^{(t')}$ to $S^{(t'+1)}$, the length of the state $S^{(t')}$ 
and the local state of site $j$.
The binomial $\left(\begin{array}{c} L-1 \\ L-n \end{array}\right)$
gives the number of possibilities for when the new particles enter 
the system.  These equations lead to Eqs.~\eqref{betaQ} and
\eqref{1-betaQ}.

If the length is $L$ at time $t+1$, there are the following
three possibilities at time $t$:
\begin{itemize}
\item[(i)] the length is $L-1$ and a new particle enters (with
  probability $\alpha$),
\item[(ii)] the length is $L$, the ($L-1$)th site is occupied, and no
  particle enters (with probability $1-\alpha$),
\item[(iii)] the length is $L+1$, the $L$th site is empty, and no
  particle enters (with probability $1-\alpha$).
\end{itemize}
Then we achieve Eqn.~\eqref{QL} for $L\in\mathbb Z_{\ge 2}$.
For $L=1$,
the case (ii) is replaced by
\begin{itemize}
\item[] the length is 1, the particle at the rightmost site does not
  leave, and no particle enters (with probability
  $(1-\alpha)(1-\beta)$).
\end{itemize}

\begin{acknowledgments}
  The authors thank Alexandru Aldea, Philip Greulich, Joachim Krug and
  Gunter M. Sch\"{u}tz for useful discussion.  This work is supported
  by Grant-in-Aid for Young Scientists ((B) 22740106) and Global COE
  Program ``Education and Research Hub for Mathematics-for-Industry.''
\end{acknowledgments}

%%%%%%%%%%%%%%%%%%%%%%%%%%%%%%%%%%%%%%%%%%%%%%%%%%%%%


\begin{thebibliography}{99}

\bibitem{RefAY}
C Arita and D Yanagisawa:
J. Stat. Phys. \textbf{141}, 829 (2010)

\bibitem{RefE}
A K Erlang:
% The theory of probabilities and telephone conversations.
Nyt. Tidsskr. Mat. Ser. B \textbf{20}, 33 (1909)

\bibitem{RefK}
D G Kendall:
% Some problems in the theory of queues.
J. Roy. Statist. Soc. Ser. B \textbf{13}(2), 151 (1951)

\bibitem{RefS} 
T L Saaty: {\em Elements of Queueing Theory With Applications},
Dover Publ. (1961)

\bibitem{RefL}
T M Liggett,
{\em Stochastic Interacting Systems:
Contact, Voter and Exclusion Processes},
Springer, New York (1999)

\bibitem{RefCSS}
D Chowdhury, L Santen and A Schadschneider:
Phys. Rep. \textbf{329}, 199 (2000)

\bibitem{RefSCN}
A Schadschneider, D Chowdhury and K Nishinari,
{\em Stochastic Transport in Complex Systems: From Molecules to Vehicles},
Elsevier Science, Amsterdam (2010)

\bibitem{RefA}
C Arita: 
Phys. Rev. E \textbf{80}, 051119 (2009)

\bibitem{RefY}
D Yanagisawa, A Tomoeda, R Jiang and K Nishinari:
JSIAM Lett. \textbf{2}, 61 (2010)

\bibitem{RefBE}
R A Blythe and M R Evans:
J. Phys. A: Math. Gen. \textbf{40}, R333 (2007)

\bibitem{RefERS}
M R Evans, N Rajewsky and E R Speer, 
J. Stat. Phys. \textbf{95}, 45--96 (1999)

\bibitem{RefKSKS}
A B Kolomeisky, G M Sch\"utz, E B Kolomeisky and J P Straley:
J. Phys. A: Math. Gen. \textbf{31} 6911 (1998)

\bibitem{RefW}
H S Wilf, {\em Generatingfunctionology}, Academic Press, San Diego
(1994)



\end{thebibliography}
\end{document}